\documentclass[10pt,a4paper,twocolumn]{aastex62}
\usepackage{graphicx,times}
\usepackage{natbib}
\usepackage{amssymb,amsmath}
\usepackage{subfigure}
\usepackage{float}

\def\beq{\begin{eqnarray}}
\def\eeq{\end{eqnarray}}

\submitjournal{ApJ}
\shorttitle{Neutrinos from hyperaccretion in collapsars}
\shortauthors{Wei et al.}
	
\begin{document}

\title{Black hole hyperaccretion in collapsars. I. MeV neutrinos}

\correspondingauthor{Tong Liu}
\email{tongliu@xmu.edu.cn}

\author{Yun-Feng Wei}
\affiliation{Department of Astronomy, Xiamen University, Xiamen, Fujian 361005, China}

\author{Tong Liu}
\affiliation{Department of Astronomy, Xiamen University, Xiamen, Fujian 361005, China}

\author{Cui-Ying Song}
\affiliation{Department of Astronomy, Xiamen University, Xiamen, Fujian 361005, China}

\begin{abstract}
As the plausible central engine of gamma-ray bursts (GRBs), a black hole (BH) hyperaccretion disk should be in a state of neutrino-dominated accretion flow (NDAF) if the accretion rate is larger than the ignition rate of an NDAF. A rotating stellar-mass BH surrounded by a hyperaccretion disk might be born in the center of a massive core collapsar. In the initial hundreds of seconds of the accretion process, the mass supply rate of the massive progenitor is generally higher than the ignition accretion rate, but the jets are generally choked in the envelope. Thus, neutrinos can be emitted from the center of a core collapsar. In this paper, we study the effects of the masses and metallicities of progenitor stars on the time-integrated spectra of electron neutrinos from NDAFs. The peak energies of the calculated spectra are approximately 10-20 MeV. The mass of a collapsar has little influence on the neutrino spectrum, and a low metallicity is beneficial to the production of low-energy ($\lesssim$ 1 MeV) neutrinos. We also investigate the differences in the electron neutrino spectra between NDAFs and proto-neutron stars. Combining with the electromagnetic counterparts and multi-messenger astronomy, one may verify the possible remnants of the core collapse of massive stars with future neutrino detectors.
\end{abstract}


\section{Introduction}

The final fate of massive stars ($> 8~M_{\odot}$) has attracted wide attention. While most massive stars end their life as a core-collapse supernova (CCSN), some do not explode (i.e., unnovae) and collapse directly to a black hole (BH) with no optical counterpart \citep[e.g.,][]{Kochanek2008}. In the collapsar model \citep[e.g.,][]{Woosely1993}, the central core of a massive star forms a BH with a hyperaccretion disk, which is widely considered to be the central engine of gamma-ray bursts (GRBs). For massive collapsars, the accretion rates are high enough to trigger neutrino cooling processes in the initial accretion phase. This hyperaccretion disk is called a neutrino-dominated accretion flow \citep[NDAF, see, e.g.,][]{Popham1999,Narayan2001,Kohri2002,Lee2005,Gu2006,Liu2007,Liu2015,Liu2017,Chen2007,Janiuk2007,Kawanaka2007,Lei2009,Xue2013,Song2016}.

The properties of NDAFs have been widely investigated over the past years \citep[for a review, see][]{Liu2017,Liu2018a,Zhang2018}. In the inner region of an NDAF, the density and temperature are so high ($\rho \sim 10^{10}-10^{13}~\rm g~cm^{-3}$, $T \sim 10^{10}-10^{11}~\rm K$) that photons are trapped. Numerous neutrinos are emitted from the surface of the disk, carrying away viscous heating energy and dissipating the gravitational energy of the BH. Neutrino cooling processes mainly include the Urca process, electron-positron annihilation, plasma decay, and the nucleon-nucleon bremsstrahlung. Therefore, three flavors of neutrinos and antineutrinos should be produced in NDAFs. The Urca process is dominant in NDAFs, so electron neutrinos and antineutrinos are the main products. It is expected that neutrinos emitted from NDAFs should have an MeV energy range, similar to those from CCSNe \citep[e.g.,][]{Liu2016}.

Based on current neutrino detectors, e.g., Super-Kamiokande (Super-K), if a supernova (SN) occurs near the Galactic center (about 10 kpc away from the Earth), about $10^{4}$ neutrino events will be detected \citep[e.g.,][]{Burrows1992,Totani1998}. Future MeV neutrino detectors could achieve better results, with the number of the events reaching approximately $10^{5}$ and $10^{6}$ for Hyper-Kamiokande (Hyper-K) and Deep-TITAND, respectively \citep[e.g.,][]{Abe2011,Kistler2011}. Although the neutrino luminosity of a typical NDAF is approximately 1-4 orders of magnitude lower than that of an SN \citep{Liu2016}, for a Galactic explosion, the number of detected neutrinos from NDAFs is still considerable. Moreover, the neutrinos from NDAFs should be detected tens of seconds later than these from SNe. Hyper-K might provide high-statistics neutrino light curves and detailed neutrino energy spectra from NDAFs and SNe at very close distances.

Considering the effects of viewing angle, BH spin, and mass accretion rate, electron neutrino/antineutrino spectra of NDAFs were calculated in \citet{Liu2016}. In this paper, we investigate time-integrated neutrino spectra of NDAFs in the collapsar scenario and compare the neutrino spectra of NDAFs with the spectrum of a proto-neutron star (PNS). According to the differences in the spectra, we expect to be able distinguish the remnant of a massive star collapse with future neutrino detectors. The paper is organized as follows. In Section 2, based on the time evolution of the mass supply rate of progenitors with different masses and metallicities, time-integrated neutrino energy spectra of NDAFs are studied. The differences in the neutrino energy spectra between NDAFs and a typical PNS are shown in Section 3, and we further discuss the possibility of distinguishing the final products of a massive star collapse with future neutrino detectors. In Section 4, we discuss the detectability of the central engines in core-collapse events. The conclusions and a discussion are presented in Section 5.

\section{Neutrinos from NDAFs in collapsars}
\subsection{Progenitor model}

\begin{figure*}[t]
\begin{minipage}{0.5\linewidth}
  \centerline{\includegraphics[angle=0,height=6cm,width=8cm]{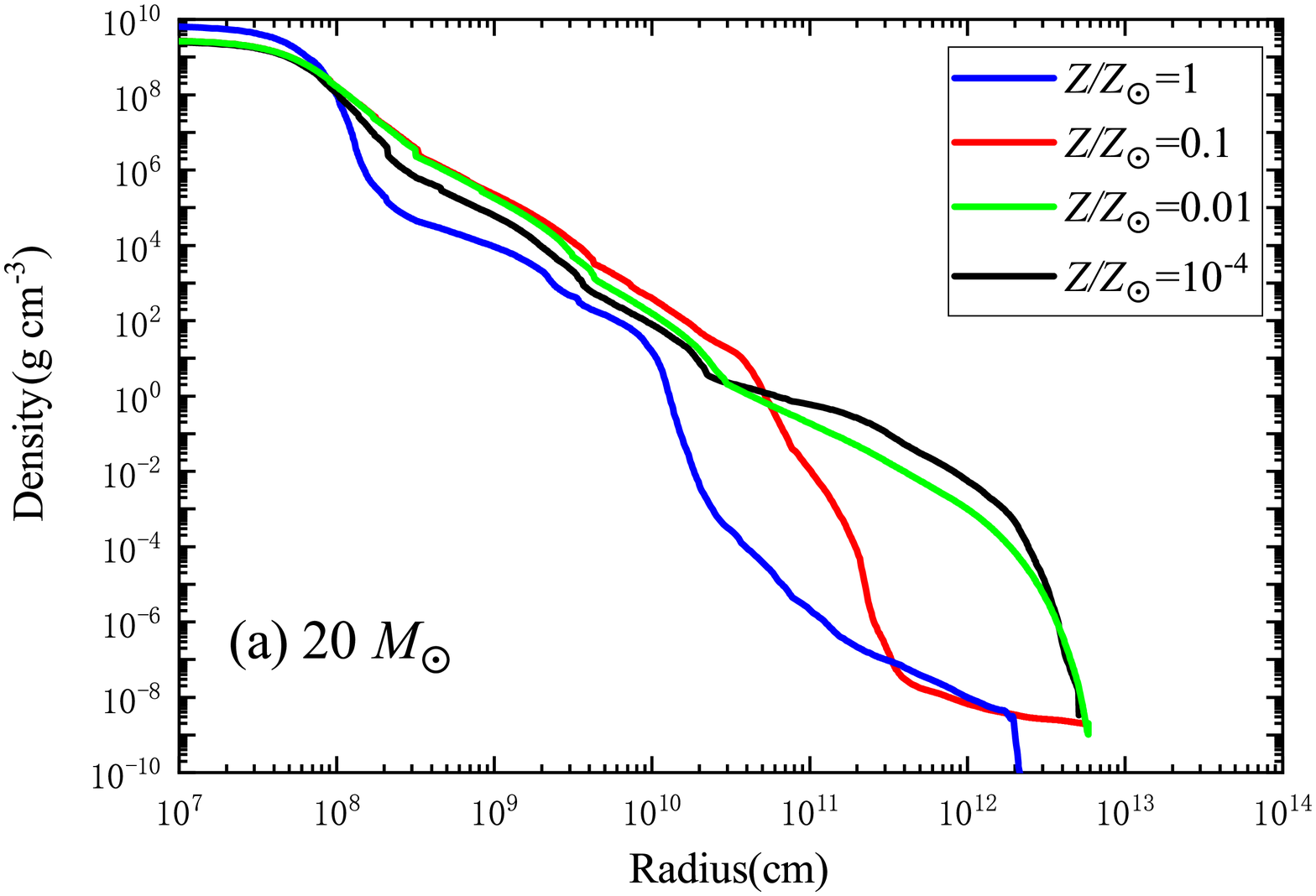}}
\end{minipage}
\hfill
\begin{minipage}{0.5\linewidth}
  \centerline{\includegraphics[angle=0,height=6cm,width=8cm]{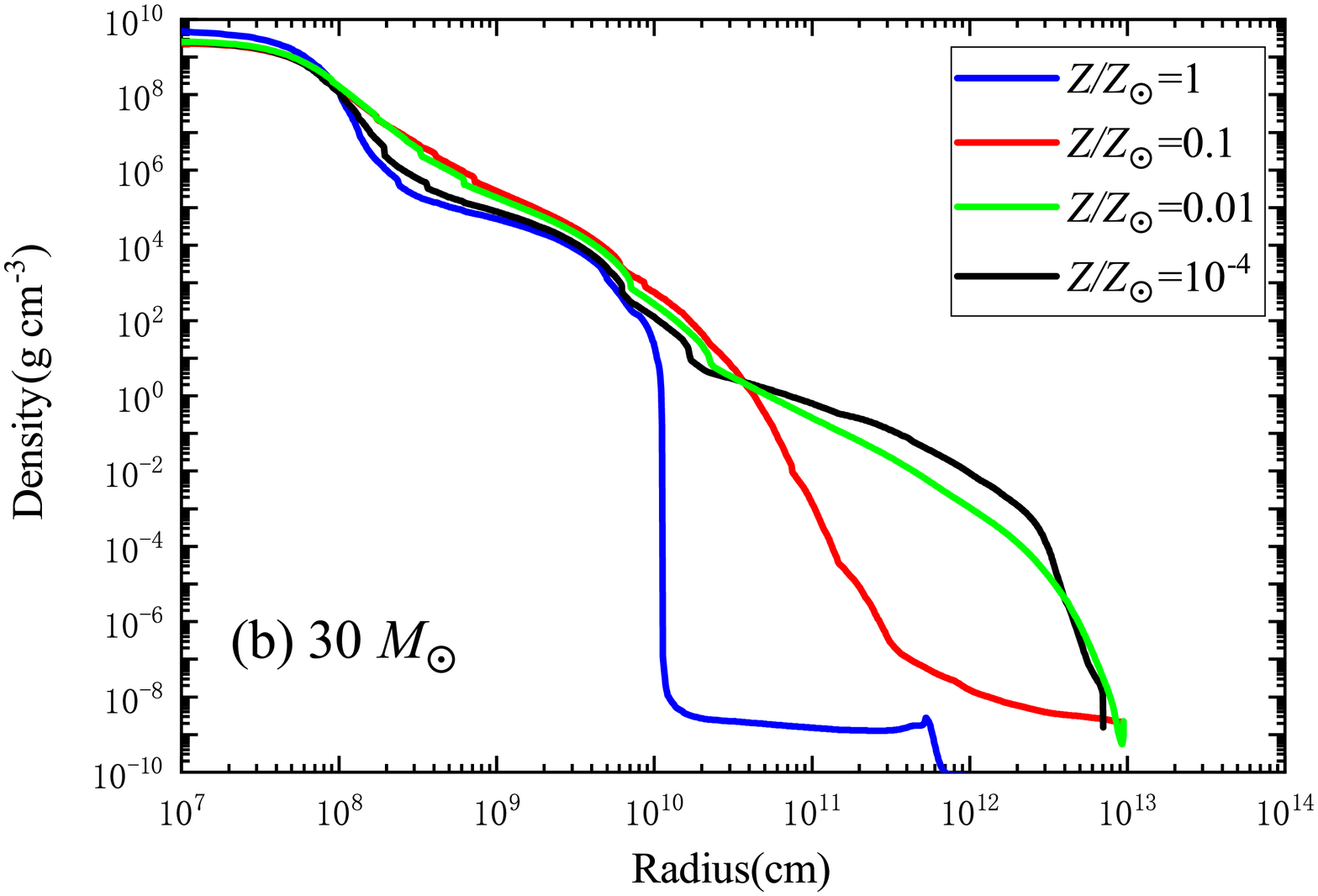}}
\end{minipage}
\vfill
\begin{minipage}{0.5\linewidth}
  \centerline{\includegraphics[angle=0,height=6cm,width=8cm]{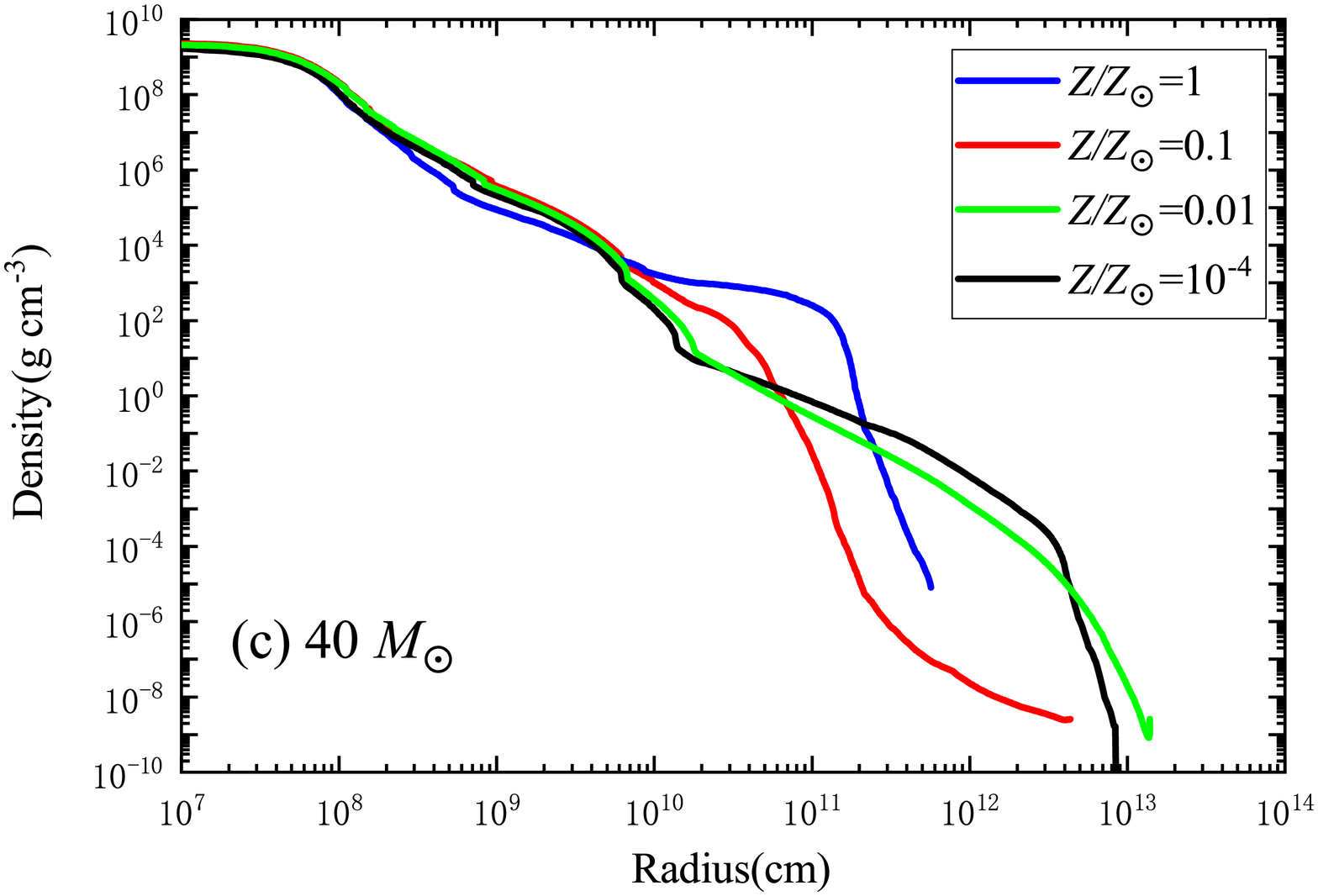}}
\end{minipage}
\hfill
\begin{minipage}{0.5\linewidth}
  \centerline{\includegraphics[angle=0,height=6cm,width=8cm]{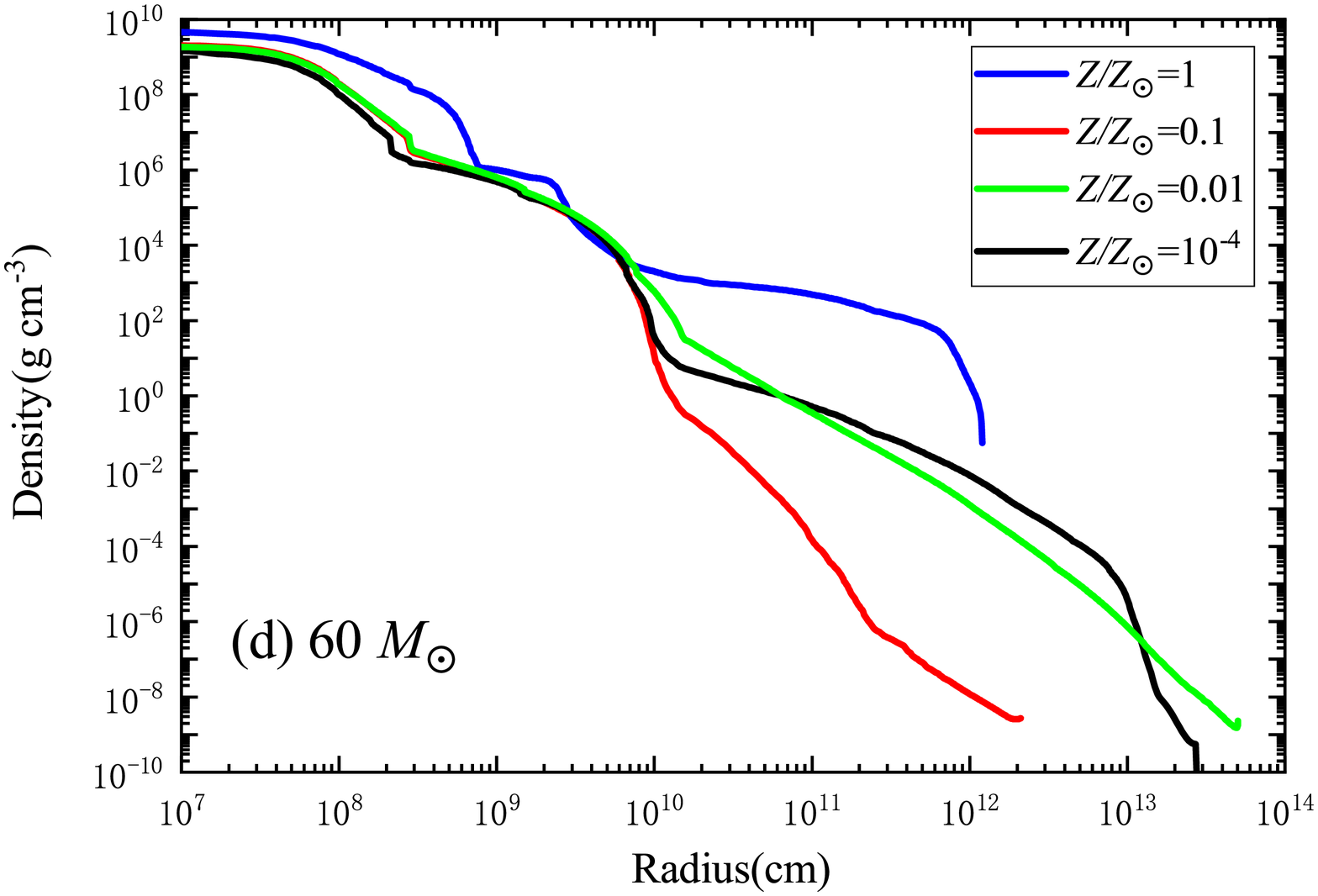}}
\end{minipage}
\caption{Density profiles of investigated progenitor stars with different masses and metallicities. The blue, red, green, and black curves correspond to metallicities of $Z/Z_{\odot}$=1, 0.1, 0.01, and $10^{-4}$, respectively.}
\end{figure*}

The pre-SN model \citep[see e.g.,][]{Woosley2002,Woosley2007,Heger2010} is adopted in our calculations. In this model, the equation of hydrostatic equilibrium of the massive stars in the pre-collapse phase can be expressed as
\beq
\frac{dP}{dR}=-\frac{G M_{R}\rho}{R^{2}},
\eeq
where $P$ is the pressure, $R$ is the radius from the center of the star, $M_R$ and $\rho$ are the mass coordinate and the density of the star, respectively. The density profiles can be obtained by the above equation.

Note that star rotation was neglected in above pre-SN models \citep[e.g.,][]{Woosley2002}. The centrifugal force was not included in the stellar hydrostatic equilibrium. This is an acceptable approximation when the ratio of the centrifugal force to the gravity remains small, for example, moderately rotating stars \citep[e.g.,][]{Woosley2006,Woosley2007}. However, for fast rotating stars with mass loss, the centrifugal force can exceed the gravity in the outer layer \citep[e.g.,][]{Chiosi1986,Maeder2012}.

We can use these density profiles to calculate the mass apply rate of progenitors \citep[e.g.,][]{Suwa2011,Woosley2012,Matsumoto2015,Liu2018,Liu2019}, i.e.,
\beq
\dot{M}_{\rm{pro}}=\frac{dM_{\rm{R}}}{dt_{\rm{ff}}}=\frac{dM_{\rm{R}}/dR}{dt_{\rm{ff}}/dt}=\frac{2M_{\rm{R}}}{t_{\rm{ff}}}(\frac{\rho}{\bar{\rho}-\rho }),
\eeq
where $\bar{\rho}=3M_{\rm{R}}/(4\pi R^{3})$ is the mean density within $R$. Of course, the mass of central BH is deducted in the density profiles \citep{Liu2018}. Here, we roughly set the accretion rate $\dot{M}$ equal to the mass apply rate \citep[e.g.,][]{Kashiyama2013,Nakauchi2013}. The free-fall timescale can be expressed as
\beq
t_{\rm{ff}}=\sqrt{\frac{3\pi}{32G\bar{\rho}}}=\frac{\pi}{2}\sqrt{\frac{R^{3}}{2GM_{\rm{R}}}}.
\eeq

In the initial hundreds of seconds of the hyperaccretion stage, the jets are generally choked in the envelope of a collapsar, so no electromagnetic counterparts of the central engine can be observed \citep[e.g.,][]{Kashiyama2013,Nakauchi2013,Liu2018,Liu2019,Song2019}. If the jets cannot break out from the envelope or the dense circumstellar medium, a jet-driven SN can form. Once the jets break out from the envelope and align with the line of sight of an observer, a long-duration GRB (LGRB) or an ultra-LGRB (ULGRB) can be observed \citep[e.g.,][]{Liu2018}

\subsection{Neutrino spectra of NDAFs}

\begin{figure*}
\begin{minipage}{0.5\linewidth}
  \centerline{\includegraphics[angle=0,height=6cm,width=8cm]{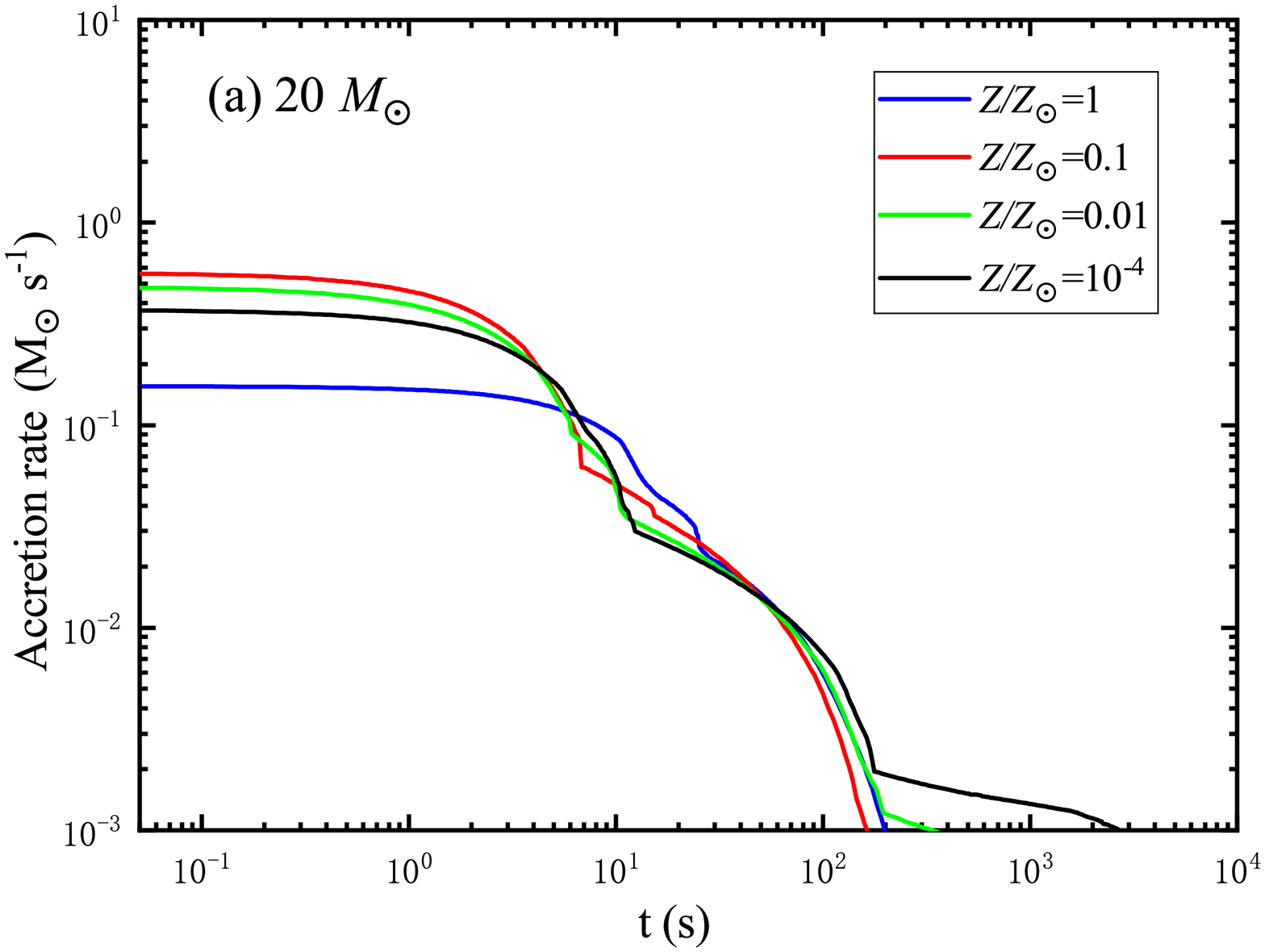}}
\end{minipage}
\hfill
\begin{minipage}{0.5\linewidth}
  \centerline{\includegraphics[angle=0,height=6cm,width=8cm]{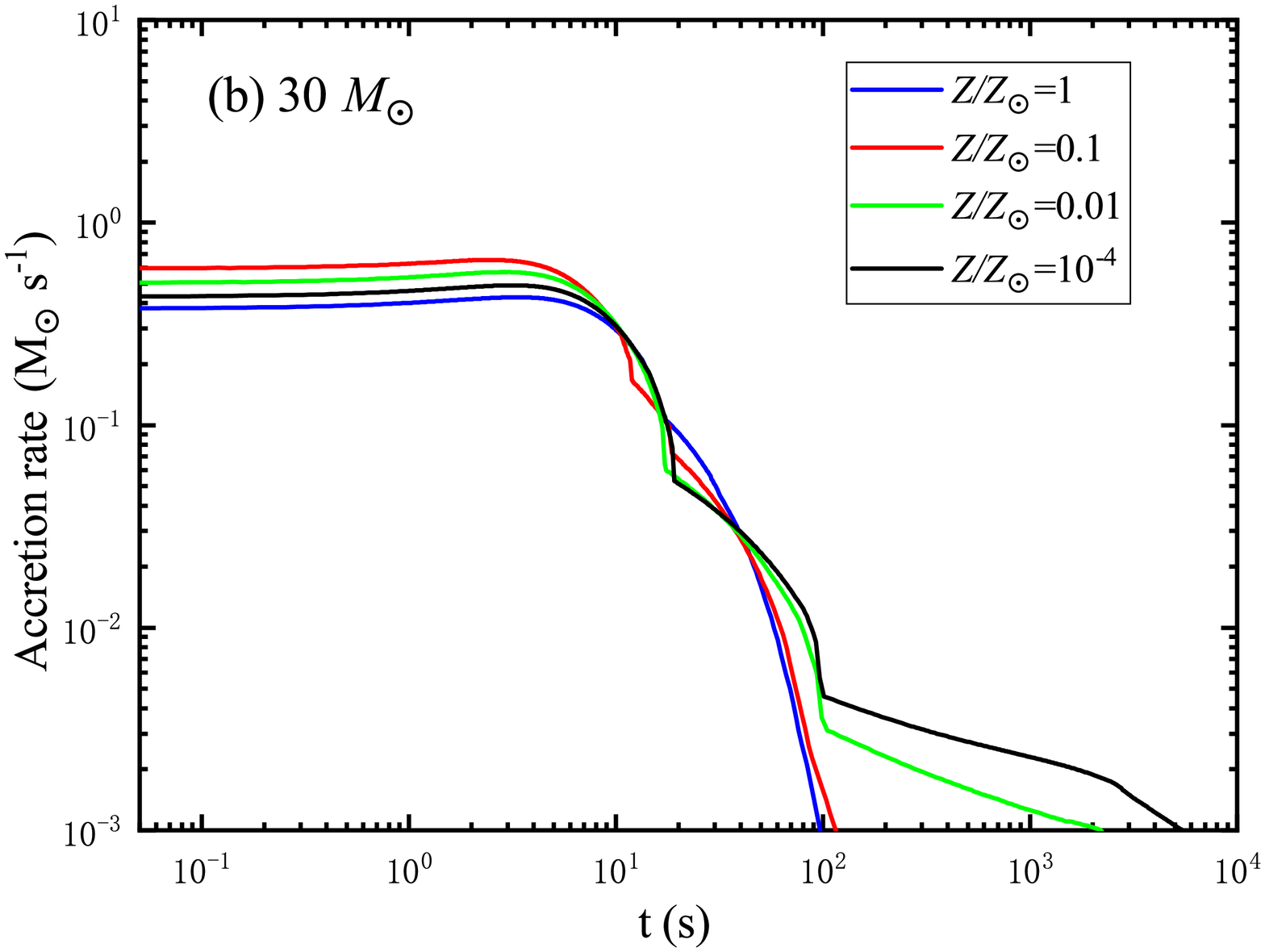}}
\end{minipage}
\vfill
\begin{minipage}{0.5\linewidth}
  \centerline{\includegraphics[angle=0,height=6cm,width=8cm]{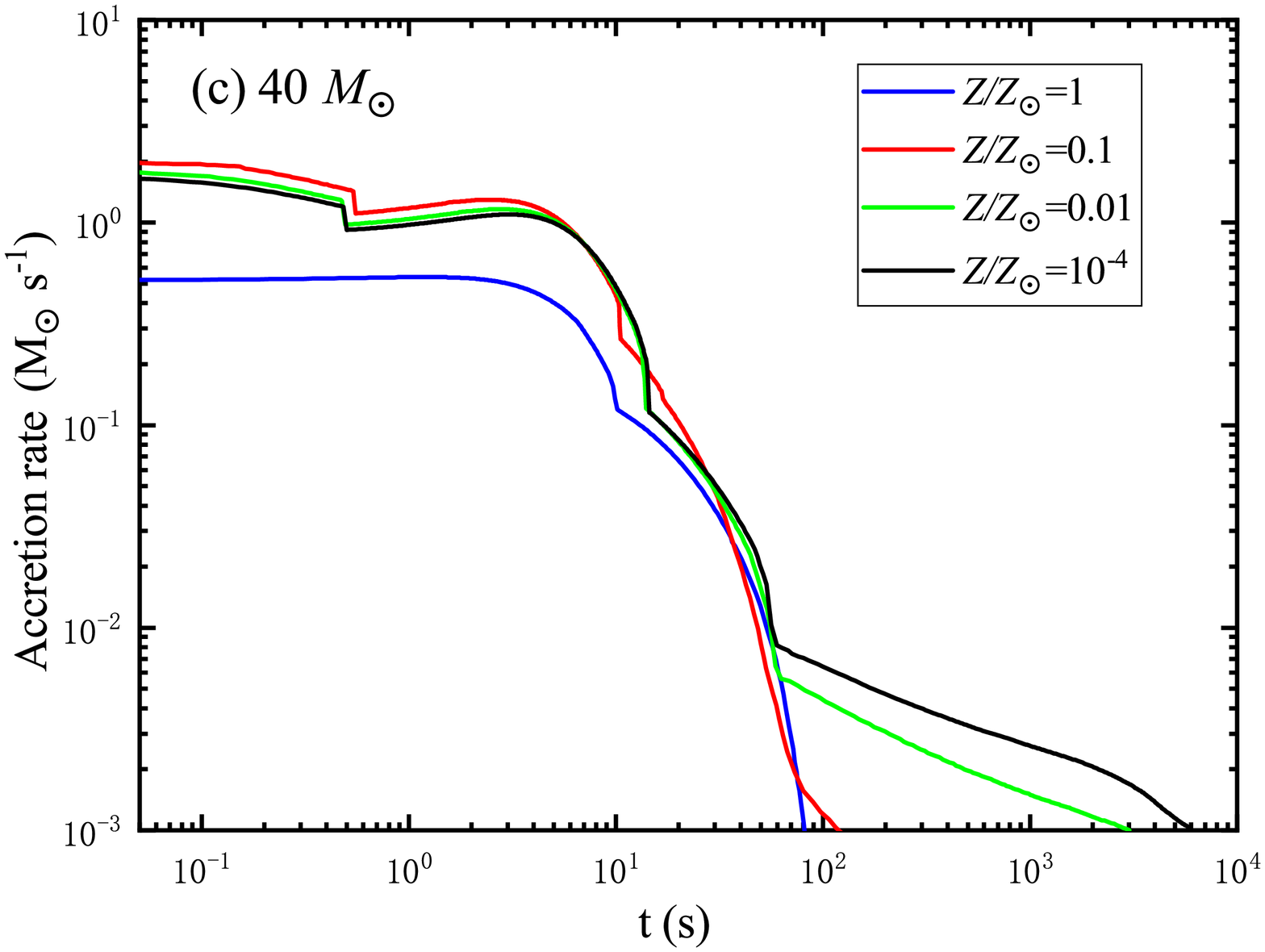}}
\end{minipage}
\hfill
\begin{minipage}{0.5\linewidth}
  \centerline{\includegraphics[angle=0,height=6cm,width=8cm]{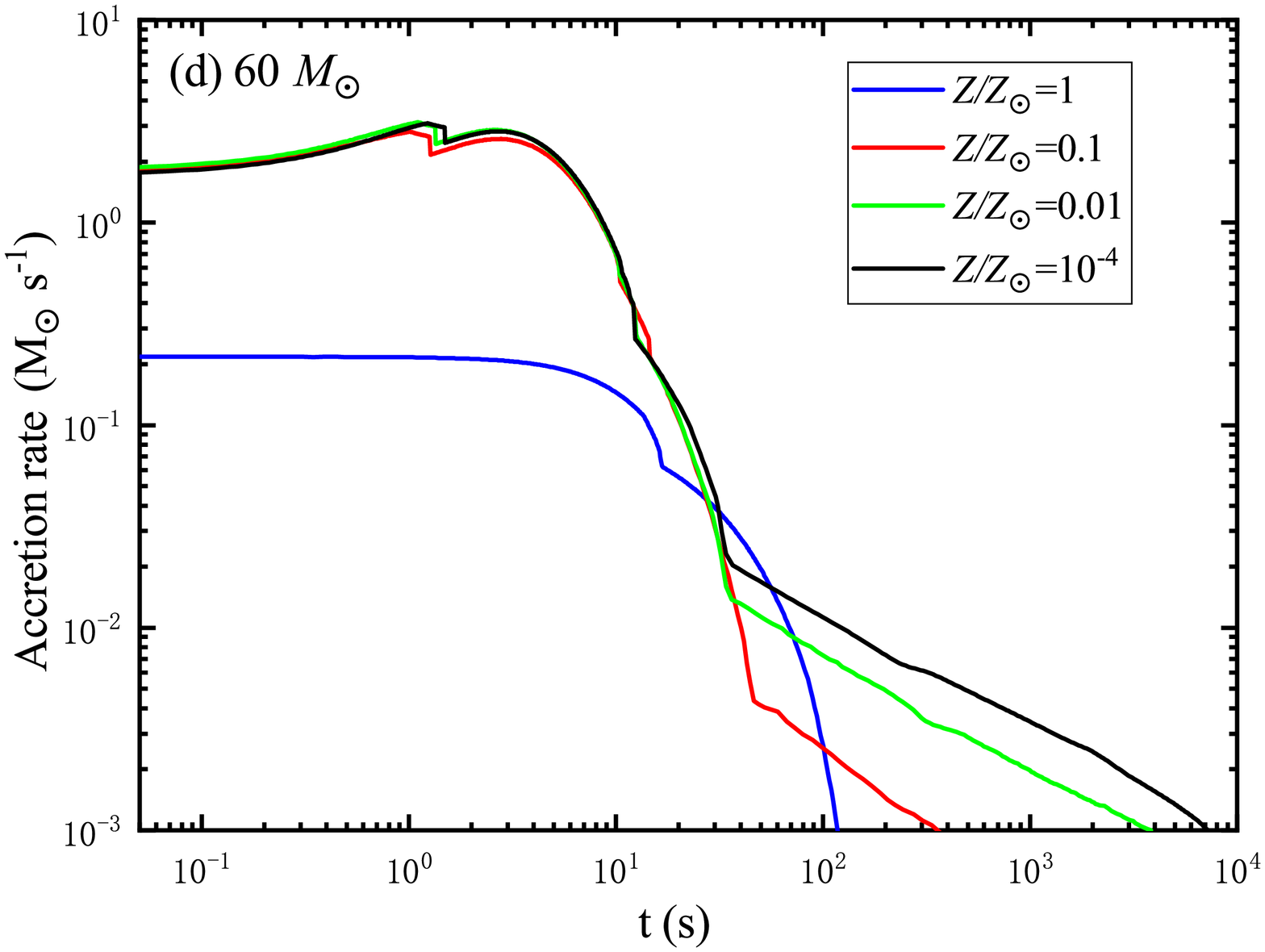}}
\end{minipage}
\caption{Time evolution of the mass accretion rate of the progenitor stars with different masses and metallicities. The blue, red, green, and black curves correspond to metallicities of $Z/Z_{\odot}$=1, 0.1, 0.01, and $10^{-4}$, respectively.}
\end{figure*}

In NDAF models, the dominant neutrino cooling process is the Urca process, and electron neutrinos and antineutrinos are the main products. There is little difference between the energy spectra of electron neutrinos and antineutrinos \citep{Liu2016}. For the $\nu$ or $\tau$ neutrinos, their yield is relatively low in NDAFs. Moreover, their detection are difficult for the current neutrino detectors \citep[e.g.,][]{Seadrow2018,Gallo2018}. Thus, we just show the electron neutrino spectra of NDAFs in this paper. Although some neutrinos and antineutrinos escape from the disk surface and annihilate above the disk, only approximately 1$\%$ of the total neutrino emission energy is consumed in this process \citep[e.g.,][]{Liu2007,Xue2013,Liu2016}. Therefore, annihilation effects can be ignored when we calculate the electron neutrino spectrum of an NDAF.

There are some characteristic radii in NDAFs \citep[e.g.,][]{Chen2007,Zalamea2011,Liu2012,Liu2017,Liu2018a,Zhang2018}, such as the ignition radius $r_{\rm ign}$. Inside the region of $r<r_{\rm ign}$, neutrino cooling processes are dominant. $r_{\rm ign}$ can be defined as the radius such that $Q_{\nu}^{-}/Q_{\rm vis}=1/2$, where $Q_{\nu}^{-}$ and $Q_{\rm vis}$ are the neutrino cooling rate and the viscous heating rate, respectively \citep[e.g.,][]{Chen2007,Liu2012}. An NDAF is ignited only when $r_{\rm ign}$ is larger than the inner radius of the disk. The corresponding mass accretion rate is $\dot{M}_{\rm ign}$, which is mainly related to the viscous parameter of the disk and the BH spin. In a collapsar, the amount of mass accreted onto the BH decreases with time \citep[e.g.,][]{Woosley2012,Liu2018}. When $\dot{M}<\dot{M}_{\rm ign} \sim 0.001~ M_\odot~\rm s^{-1}$, the disk is no longer called an NDAF, and neutrino emission is ignored. Here, we can define the time when $\dot{M}$ decreases to $\dot{M}_{\rm ign}$ as $t_{\rm ign}$, which is typically hundreds of seconds.

Based on the solutions of \citet{Xue2013}, \citet{Liu2016} derived fitting formulae for the mean cooling rate due to electron neutrino losses, $Q_{\nu_{\rm{e}}}$, in units of $\rm erg~cm^{-2}~s^{-1}$, and the temperature of the disk $T$, in units of $\rm K$, as a function of the mean BH spin parameter, the accretion rate, and the radius, i.e.,
\beq
\log Q_{\nu_{\rm{e}}}=39.78+0.15a_{*}+1.19\log \dot{m}-3.46\log r,
\eeq
\beq
\log T=11.09+0.10a_{*}+0.20\log \dot{m}-0.59\log r,
\eeq
where $a_{\ast}(0\leq a_{\ast}\leq 1)$ is the dimensionless BH spin parameter, and $\dot{m}=\dot{M}/M_\odot~\rm s^{-1}$ and $r=R/R_g$ are the dimensionless mass accretion rate and radius, respectively. $R_g=2GM_{\rm{BH}}/c^2$ is the Schwarzschild radius, and the mass and spin of the BH are adopted as $M_{\rm{BH}}=3M_{\odot}$ and $a_{*}=0.9$ in our calculations.

Neutrinos are mainly emitted from the inner region of an NDAF. Therefore, the observed spectra are affected by general relativistic effects due to the strong gravitational field of the BH, such as frame-dragging, Doppler boosting, gravitational redshifts, and the bending of neutrino trajectories. We use ray-tracing methods \citep[e.g.,][]{Fanton1997,Li2005} to calculate neutrino propagation in a manner similar to photon propagation near an accreting BH and then obtain the observed neutrino spectrum. We assume that the neutrino source is located in the equatorial plane and that the neutrino emission from a certain radius is isotropic, i.e., $i_{\rm em}=\pi/2$, and the accretion disk is treated as a Keplerian disk. The shading effect due to the thickness of the disk is neglected. Neutrino trajectories originating at the emitter must satisfy the geodesic equation \citep[e.g.,][]{Carter1968}, i.e.,
\beq
\pm \int_{R_{\rm em}}^{\infty }\frac{dR}{\sqrt{l(R)}}=\pm \int_{i_{\rm em}}^{i_{\rm obs} }\frac{di}{\sqrt{I(i)}}.
\eeq

\begin{figure*}
\begin{minipage}{0.5\linewidth}
  \centerline{\includegraphics[height=6cm,width=8cm]{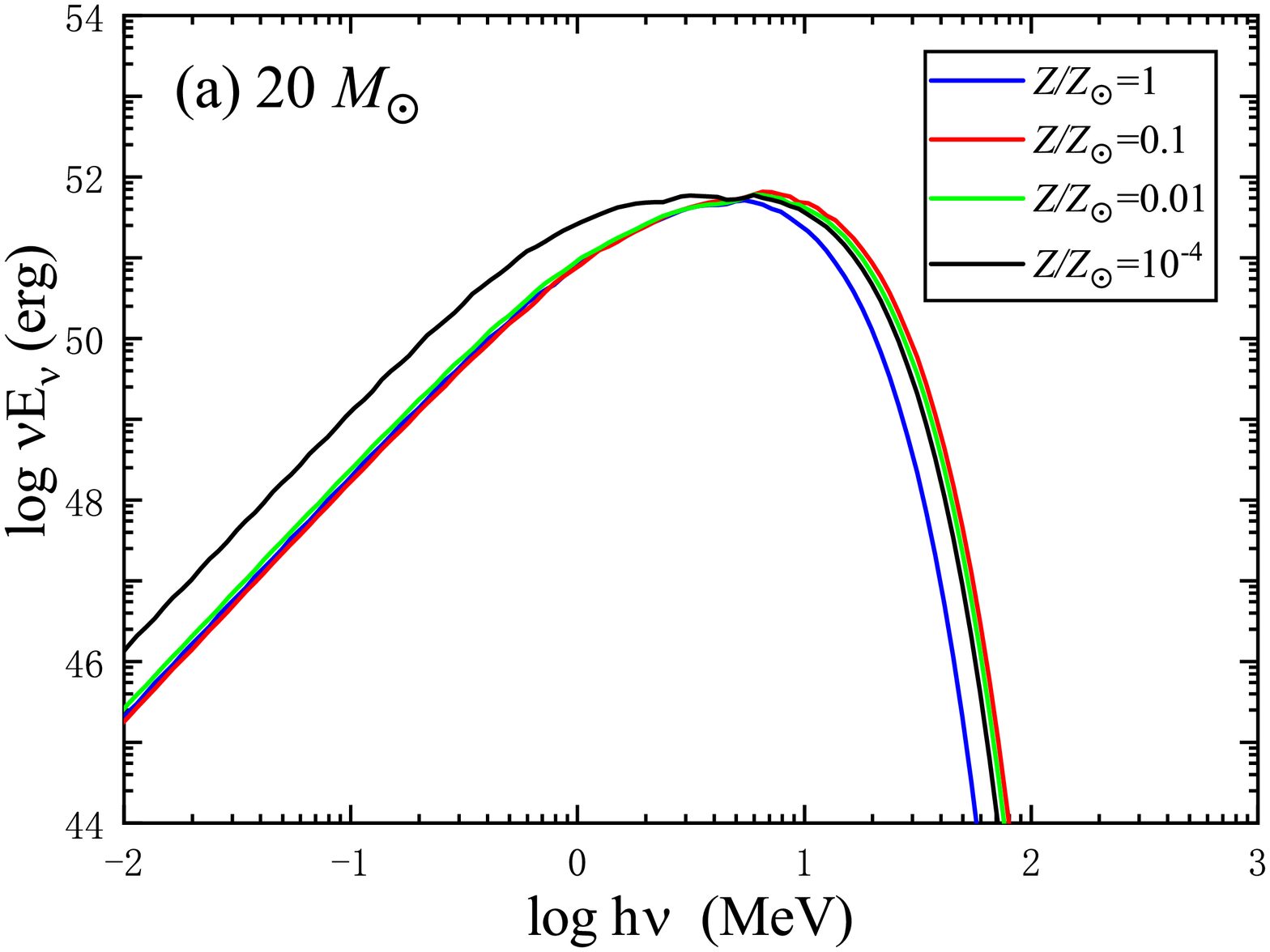}}
\end{minipage}
\hfill
\begin{minipage}{0.5\linewidth}
  \centerline{\includegraphics[height=6cm,width=8cm]{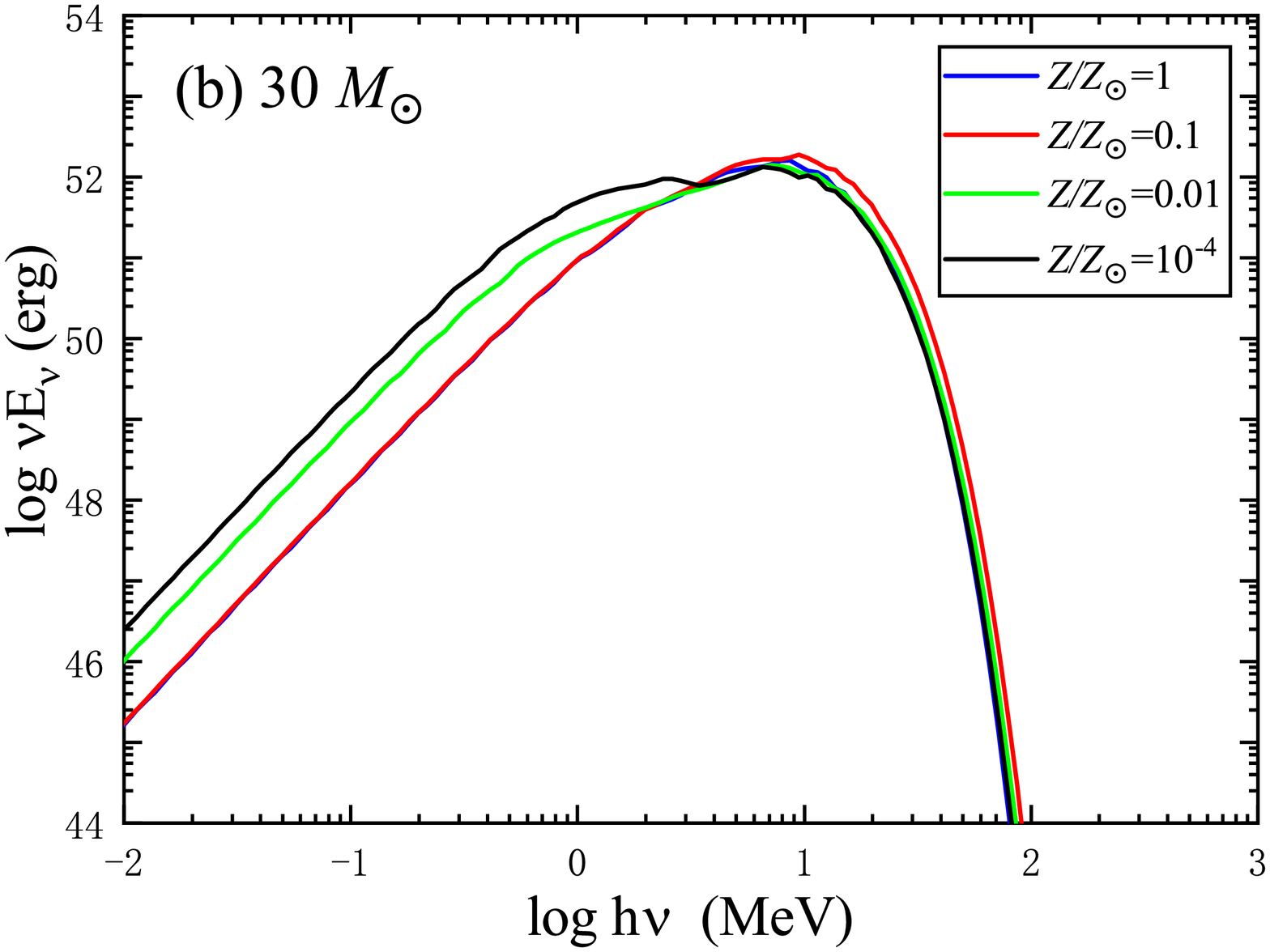}}
\end{minipage}
\vfill
\begin{minipage}{0.5\linewidth}
  \centerline{\includegraphics[height=6cm,width=8cm]{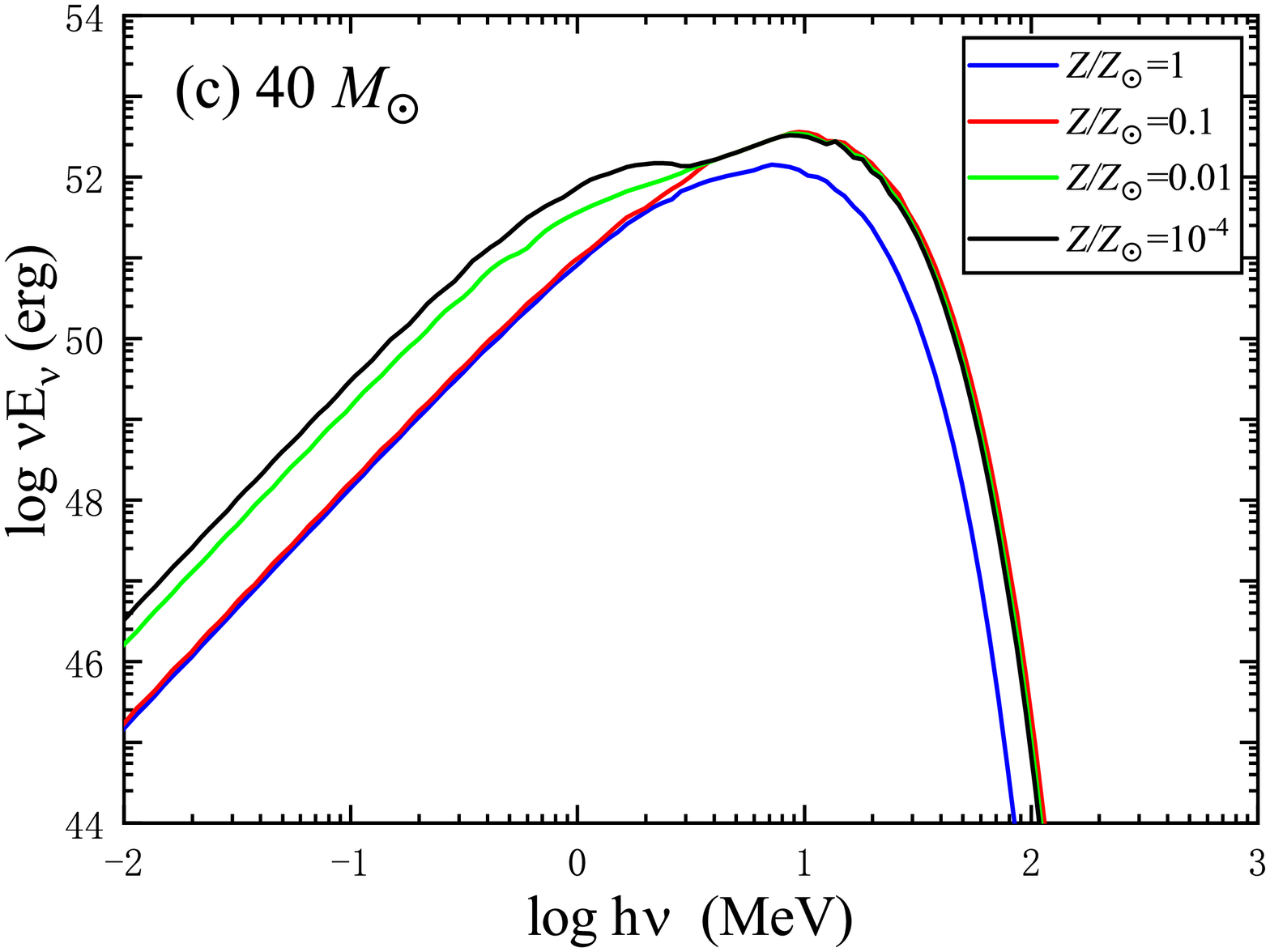}}
\end{minipage}
\hfill
\begin{minipage}{0.5\linewidth}
  \centerline{\includegraphics[height=6cm,width=8cm]{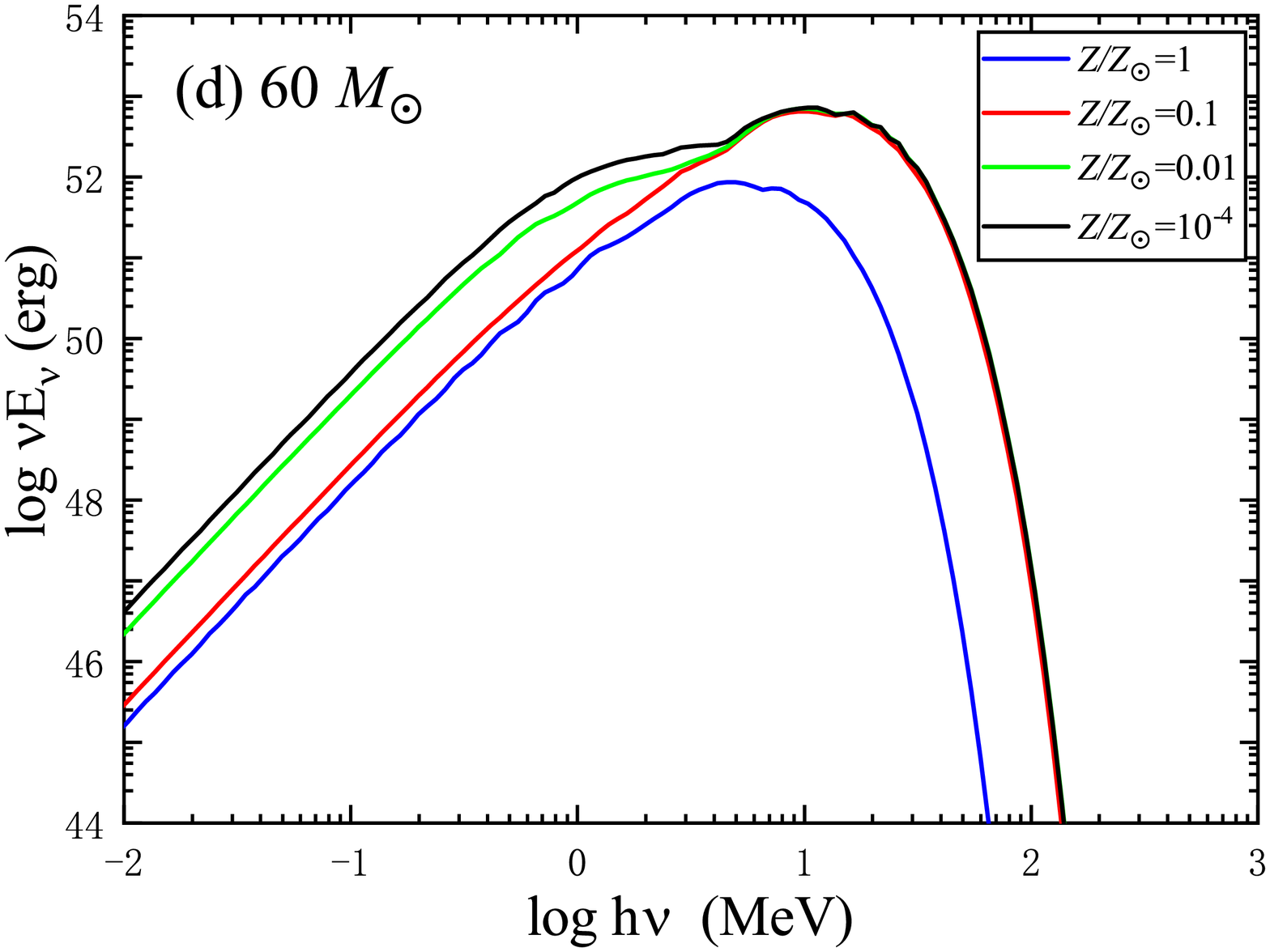}}
\end{minipage}
\caption{Electron neutrino spectra of NDAFs with different masses and metallicities. The blue, red, green, and black curves correspond to progenitor star metallicities of $Z/Z_{\odot}$=1, 0.1, 0.01, and $10^{-4}$, respectively.}
\end{figure*}

The above equation is satisfied by the null geodesic that reaches the observer from the source. Based on the null geodesic equation, for each pixel of the observed image, the position of the emitter on the disk can be traced. By investigating the corresponding velocity and gravitational potential of the emission location, we can calculate the energy shift of a neutrino emitted from this position. The energy shift factor is defined as $g \equiv E_{\rm obs} / E_{\rm em}$. Integrating over all the pixels, the total observed flux distribution can be calculated. i.e.,
\beq
F_{{E}_{\rm{obs}}}=\int_{\rm image}g^{3}I_{E_{\rm{em}}}d\Omega _{\rm{obs}},
\eeq
where $E_{\rm{obs}}$ is the observed neutrino energy, $E_{\rm{em}}$ is the neutrino emission energy from the local disk, and $\Omega _{\rm{obs}}$ is the solid angle of the disk image to the observer. $I_{E_{\rm{em}}}$ is the local emissivity, which can be obtained by the cooling rate $Q_{\nu_{\rm{e}}}$ as
\beq
I_{E_{\rm{em}}}=Q_{\rm{\nu }}\frac{F_{E_{\rm{em}}}}{\int F_{E_{\rm{em}}}dE_{\rm{em}}},
\eeq
where $F_{E_{\rm{em}}}=E_{\rm{em}}^{2}/[\exp (E_{\rm{em}}/kT-\eta )+1]$ is the unnormalized Fermi-Dirac spectrum \citep[e.g.,][]{Rauch1994,Fanton1997,Li2005,Liu2016}.

Then, the luminosity distribution is given by
\beq
L_{\rm{\nu} }=4\pi D_{\rm{L}}^{2} F_{{E}_{\rm{obs}}},
\eeq
where $D_{\rm{L}}$ is the luminosity distance in the standard $\rm{\Lambda CDM}$ cosmology model with $\Omega _{M}=0.27, \Omega _{\Lambda }=0.73$, and $H_{0}=71\,\rm{km} \rm{s}^{-1} \rm{Mpc}^{-1}$ \citep[e.g.,][]{Liu2015}.

The time-integrated neutrino energy spectrum is given by
\beq
E_{\rm{\nu }}=\int_{t_{0}}^{t_{\rm{ign}}}L_{\rm{\nu}} dt,
\eeq
where $t_{\rm{0}}$ is the initial accretion time. Here, the viewing angle is adopted as $i_{\rm obs}=1^\circ$.

\subsection{Results}

We select progenitor masses of $M_{\rm{pro}}/M_{\odot}$=20, 30, 40, and 60 and metallicities of $Z/Z_{\odot}$=1, 0.1, 0.01, and $10^{-4}$, where $Z_{\odot}$ is the metallicity of the Sun, to investigate the effects of mass and metallicity on the neutrino spectrum. The density profiles of the investigated progenitors stars are shown in Figure 1.

Figure 2 shows the time evolution of the mass accretion rate of progenitor stars with different masses and metallicities. The blue, red, green, and black curves correspond to progenitor star metallicities of $Z/Z_{\odot}$=1, 0.1, 0.01, and $10^{-4}$, respectively. The significant difference of $\dot{M}$ for the different progenitor stars mainly come from the different density profiles of the stars shown in Figure 1. One can notice that the mass accretion rates are all approximately 1 $M_{\odot}~\rm{s}^{-1}$ in the initial accretion stage and decrease to the ignition rate in hundreds of seconds. We also found that the decrease in the timescale of the mass accretion rates from their initial values to $\dot{M}_{\rm{ign}}$ generally depends on the metallicity, which means that the metallicity of a progenitor star plays an important role in influencing the duration of neutrino emission in a collapsar. Moreover, the progenitor mass has little influence on the mass accretion rate.

Time-integrated electron neutrino spectra of NDAFs in the center of collapsars are displayed in Figure 3. The blue, red, green, and black curves correspond to metallicities of $Z/Z_{\odot}$=1, 0.1, 0.01, and $10^{-4}$, respectively. First, there is almost no influence of total mass of the star on the neutrino spectra. Actually, the accretion rate is computed from the free-fall time scale, and depends on the mass enclosed within inner parts of the star, while the total mass is determined by the extended envelope outside this radius. Second, we found that the peaks of the spectra all occur at approximately 10-20 MeV. For NDAF models, a high accretion rate corresponds to a high disk temperature, which mainly produces high-energy neutrinos, i.e., most high-energy neutrinos are emitted in the early hyperaccretion stage with a high accretion rate. In the low-energy range of the spectra, the amplitudes of the spectral lines increase with decreasing metallicity, because a lower metallicity corresponds to neutrino emission with a longer timescale. In the late accretion stage of an NDAF, the disk mainly emits low-energy neutrinos, which can increase the amplitudes of the spectral lines of a progenitor with a low metallicity. In the high-energy range of the energy spectra, the amplitudes of the spectral lines depend on the initial accretion rate of the NDAF, which are determined by the mass and metallicity of the progenitor star.

It is worth noting that the various rotation profiles imposed on the collapsing stars may significantly change the evolution of GRBs \citep{Janiuk2008}. Thus, especially for the fast-rotating progenitor stars, the rotation might dramatically affect on the shape of the neutrino spectra when the star collapse to trigger the BH hyperaccretion.

\section{Comparisons with neutrino spectra of PNSs}

Besides BHs, neutron stars (NSs) could also be compact remnants of massive core collapsars from fast-rotating stars or stars that quickly lose their envelopes through binary interactions \citep[e.g.,][]{Podsiadlowski2004a}. LGRBs or ULGRBs and the associated super-luminous SNe have been investigated in the scenario of a magnetar \citep[a rapidly spinning, supramassive, strongly magnetized NS, see, e.g.,][]{Duncan1992,Usov1992,Dai1998,Zhang2001,Dai2006,Zhang2018} located in the center of a collapsar \citep[e.g.,][]{Metzger2011,Metzger2015}.

\begin{figure}
\centering
\includegraphics[angle=0,scale=0.35]{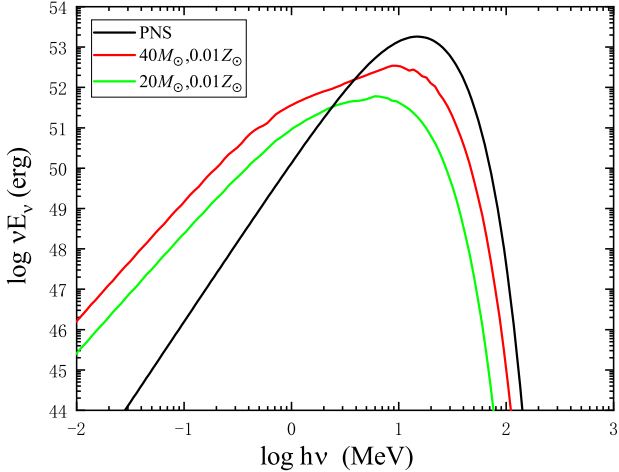}
\caption{Time-integrated electron neutrino spectra of a PNS and NDAFs in the center of a collapsar. The black curve corresponds to a PNS, and the green and red lines correspond to NDAFs in the center of progenitor stars with ($M_{\rm{pro}}/M_{\odot}$, $Z/Z_{\odot}$) = (20, 0.01) and (40, 0.01), respectively.}
\end{figure}

A newborn NS is proton-rich and contains a large number of neutrinos and degenerate electrons \citep[e.g.,][]{Janka2007}. Neutrino emission plays a key role in deleptonization and cooling of a hot PNS during the Kelvin-Helmholtz epoch. It takes a hot PNS tens of seconds to cool to a cold and deleptonized NS \citep[e.g.,][]{Pons1999}. It is well known that neutrinos are nearly massless  particles and are only affected by the weak interaction with extremely small scattering cross-sections. Here, neutrino flavor oscillations is neglected. However, the matter of a PNS is quite opaque at these high temperatures and densities, and neutrinos produced in a hot PNS are trapped and escape by diffusion \citep[e.g.,][]{Burrows1986,Burrows1990}. The main mechanisms contributing to the opacity are neutral current scattering reactions and charged current absorption reactions \citep[e.g.,][]{Burrows1986,Burrows1990,Yuan2017}. Regardless of the mechanism, opaque neutrino emission from PNSs has the form of blackbody emission \citep[e.g.,][]{Janka1989a,Janka1989b}. The neutrino spectrum conforms to the Fermi-Dirac energy distribution, which is determined by the neutrinospheric temperature. Many studies in the literature have performed numerical calculations and simulations of the cooling of PNSs \citep[e.g.,][]{Burrows1986,Pons1999,Hudepohl2010}.

Here, we adopt part of the simulation results in \citet{Pons1999}, i.e., the cooling of a $\sim 1.8~M_{\odot}$ PNS in the GM3np model, to describe the time-integrated electron neutrino spectrum of a PNS, which is shown in Figure 4. As a comparison, neutrino spectra of NDAFs corresponding to progenitor stars with ($M_{\rm{pro}}/M_{\odot}$, $Z/Z_{\odot}$) = (20, 0.01) and (40, 0.01) are also displayed in the figure. The main differences in the spectra occur in the low-energy region. The spectra of the NDAFs definitely exhibit a more gentle ascent than the PNS spectrum. Above several MeV, the amplitudes of the PNS spectrum are about one order of magnitude higher than those of the NDAF spectra.

In the collapsar model, the initial mass accretion rate is about $1~M_{\odot}~\rm s^{-1}$; thus, the neutrino luminosity of an NDAF is lower than that of a PNS. However, for a high mass accretion rate (e.g., $9~M_{\odot}~\rm s^{-1}$) in the scenario of compact object mergers, the neutrino luminosity of an NDAF would be higher than that of a PNS \citep[e.g.,][]{Liu2017,Schilbach2018}. One can expect that neutrino spectra of the Galactic explosions from the central engines of collapsars or mergers might be captured by future neutrino detectors, allowing the elucidation of the kind of engine in the center.

In some core-collapse simulations, a fraction of the total emitted neutrinos are emitted in a hyperaccretion phase lasting about several hundred milliseconds before PNS cooling \citep[e.g.,][]{Nakazato2013,Hudepohl2010}, while most of the neutrinos are emitted later during the subsequent tens of seconds of PNS cooling and deleptonization \citep[e.g.,][]{Burrows1986}. Thus, a high-energy tail will be superimposed on the spectrum \citep{Nakazato2013}. Nonetheless, the neutrino energies from the PNS are still higher than those of an NDAF in the high-energy band and can be distinguished.

In addition, it is possible that the central core forms an NS first and then collapses to a BH surrounded by an accretion disk due to the fallback process \citep[e.g.,][]{Woosley2000,Nagataki2002}. In this case, the neutrinos from a collapsar are powered by two sources, i.e., the PNS and the hyperaccretion disk. Since the timescale over which an NS collapses to a BH is unknown, we can use a neutrino light curve to test for the existence of this process. Because the neutrino luminosity of a PNS is higher than that of an NDAF, the neutrino flux would show a rapid decrease when a PNS collapses to a BH \citep[e.g.,][]{Burrows1988,Liu2016}.

\section{Detection}

Twenty-five neutrinos from SN1987A in the Large Magellanic Cloud (LMC) were detected through inverse beta decay with the Kamiokande \citep{Hirata1987}, IMB \citep{Bionta1987} and Baksan detectors \citep{Alexeyev1988}. The discovery initiated a new era of neutrino astrophysics.

Since 1987, the capability for worldwide SN neutrino detection has improved quickly. For an SN with a total neutrino energy of $\sim 3\times 10^{53}$ ergs at a distance of about 10 kpc, 170,000-260,000 neutrino events are expected to be detected by the Hyper-K detector in the near future. In the case of the LMC where SN1987A is located, 7,000-10,000 neutrino events can be expected \citep{Abe2011}. Meanwhile, other projects on MeV neutrino detection, such as JUNO (Jiangmen Underground Neutrino Observatory), Super-K, and DUNE (Deep Underground Neutrino Experiment), declared that they can also detect neutrinos with energies of many thousands of MeV from a nearby core-collapsar event \citep[e.g.,][]{Seadrow2018}.

Although low-energy ($\lesssim 1~\rm MeV$) neutrino detection is difficult, the future liquid-scintillator detector LENA (Low Energy Neutrino Astronomy) will have a good performance in the low-energy energy band \citep{Wurm2012}. LENA may provide information on sub-MeV neutrinos from the Galactic NDAFs and PNSs.

\citet{Liu2016} presented the detection rate of NDAFs in detail. If one takes the event rate of the SN Ib/c \citep[e.g.,][]{Podsiadlowski2004,Sun2015} as an optimistic event rate for NDAFs, according to estimate the SN events of the major galaxies in the Local Group, the expected detection rate for NDAFs in the Local Group is 1-3 per century for the Hyper-K detector \citep{Liu2016}. However, only several neutrinos can be detected by Hyper-K at this distance. It is impossible to obtain detailed neutrino light curves and spectra of the central engines. We should focus on the closer neutrino bursts or develop more powerful detectors. According to the distribution of possible massive progenitors in the Milky Way, the most likely distance of the next core-collapsar event from the Sun is in the range of 12-15 kpc in the Galactic plane \citep[e.g.,][]{Tammann1994,Mirizzi2006,Scholberg2012}. For collapsars at a such distance, we might optimistically detect $\sim$ 1,000 neutrino events from the central engine by LENA (estimation by the data in \citet{Wurm2012}). Due to compare with the theoretical low-energy band of the time-integrated neutrino energy spectra resulted from NDAFs or PNSs, thereby one may roughly distinguish the nature of the remnants of the death of massive stars.

The current state is that the event rate of Galactic CCSNe and the detection rate of neutrinos are very low. In any case, it is very hard to achieve direct detection of the MeV neutrinos from the central engine of nearby CCSNe at the moment. The joint observations of the electromagnetic counterparts and gravitational waves (GWs) of the central engine should be more likely to constrain the nature of the remnants \citep[e.g.,][]{Liu2017,Liu2018}. In addition, future neutrino detectors with a low energy threshold, high signal rate, and good energy resolution are expected to be capable of providing the high-statistics light curve and spectrum of a neutrino burst from the core collapse of a massive star. Then, explosion mechanisms and neutrino physics, such as mass and its hierarchy, mixing and oscillation, could be intensively investigated \citep[e.g.,][]{Scholberg2012}.

\section{Conclusions and discussion}

We have investigated the effects of the mass and metallicity of a progenitor star on the  time-integrated spectrum of electron neutrinos from an NDAF. We found that the mass of a collapsar has little influence on the spectrum, and a low metallicity is beneficial to the production of low-energy ($\lesssim$ 1 MeV) neutrinos. The electron neutrino spectra of NDAFs and PNSs can be distinguished. One can expect to verify the remnants in the center of core-collapsars in the era of multi-messenger astronomy. The future neutrino detectors might have the capacity to provide high-statistics data on neutrinos and detailed neutrino spectra and light curves of nearby collapsars to help us verify the possible remnants of the core collapse of massive stars.

In our above calculations, we have not taken into count neutrino oscillations, which are expected to play an important role in collapsars \citep[e.g.,][]{Malkus2012}. Neutrinos emitted by the core would pass through the stellar envelope of the progenitor star. In this region, flavor transformations have a more significant influence on the flux of electron neutrinos than on the flux of electron antineutrinos \citep[e.g.,][]{Friedland2010,Duan2011}. However, in vacuum, flavor transformations have similar influences on the fluxes of electron neutrinos and electron antineutrinos. For NDAFs, which have similar flavor distribution and physical conditions to SNe, neutrino oscillations will change the neutrino spectra and reduce the neutrino flux of NDAFs by at most a factor of 2-3 \citep[e.g.,][]{Cherry2012,Friedland2010,Duan2011,Liu2016}. Neutrino oscillations can be addressed in postprocessing to calculate the final neutrino signal that reaches detectors on Earth \citep[e.g.,][]{Kotake2006}.

Furthermore, another way to distinguish the remnants of massive stars is GW detection. In the Local Group, GWs caused by anisotropic neutrino emission \citep[e.g.,][]{Epstein1978,Suwa2009,Kotake2012,Liu2017a} or jet precession \citep[e.g.,][]{Romero2010,Sun2012,Liu2017a} from NDAFs could be detected by Advanced LIGO, DECIGO/BBO, and ultimate-DECIGO. For the case of anisotropic neutrino emission, only BH hyperaccretion in the state of ignited NDAF can power GWs \citep{Liu2017a}. In future work, we will discuss GWs from NDAFs in a core-collapsar scenario. Future neutrino and GW detections would provide more information on the central engines of the collapse of massive stars, and promote the update of the collapsar model including the rotation, magnetic field, wind, nucleosynthesis, relativistic effects and so on.

\acknowledgments
We thank the anonymous referee for very useful suggestions and comments. We appreciate Prof. A. Heger who provides pre-SN data. This work was supported by the National Natural Science Foundation of China under grant 11822304.

\end{document}